\documentclass[12pt,english]{article}
\usepackage[utf8]{inputenc}
\usepackage[english]{babel}
\usepackage{float}
\usepackage{booktabs,longtable}
\usepackage{amsmath}
\usepackage{amssymb,amsthm,amsfonts}
\usepackage{enumerate}
\usepackage{tikz}
\usepackage[authoryear,round]{natbib}
\usepackage[hyphens]{url}
\usepackage{hyperref}
\usepackage[a4paper,body={16cm,23.7cm}]{geometry}

\hypersetup{colorlinks=true, linkcolor=cyan, citecolor=cyan, urlcolor=black, breaklinks=true}
\newtheorem{theorem}{Theorem}

\newtheorem{example}{Example}

\begin{document}

\title{\textbf{Fair allocation of riparian water rights}\thanks{We thank Atila Abdulkadiro\u{g}lu, Erik Ansink, Sylvain B\'{e}al, Rene van den Brink, Jens Gudmundsson, Takayuki Oishi, Emel Öztürk, William Thomson, Hans Peter Weikard and participants at the 2023 SAET Meeting, the SWELL online seminar, the 2024 Alicante Workshop on Collective Decisions and Economic Design, and the 19th European Meeting on Game Theory for helpful comments and suggestions. Financial support from the Spanish Agencia Estatal de Investigaci\'{o}n (AEI) through grants PID2020-115011GB-I00 and PID2020-114309GB-I00 is gratefully acknowledged.
}}
\author{\textbf{Ricardo Mart\'{\i}nez}\thanks{Universidad de Granada.}$\qquad$ \textbf{Juan
D. Moreno-Ternero}\thanks{Universidad Pablo
de Olavide.} \quad } \maketitle

\begin{abstract}
We take an axiomatic approach to the allocation of riparian water rights. We formalize ethical or structural properties as axioms of allocation rules. 
We show that several combinations of these axioms characterize focal rules implementing 
the principle of \textit{Territorial Integration of all Basin States} in various forms. One of them connects to the \textit{Shapley value}, the long-standing centerpiece of cooperative game theory. The others offer natural compromises between the polar principles of \textit{Absolute Territorial Sovereignty} and \textit{Unlimited Territorial Integrity}. We complete our study with an empirical application to the allocation of riparian water rights in the Nile River. 
\end{abstract}

\medskip{}

\medskip{}  \medskip{}

\newpage


\section{Introduction}
Numerous alarming questions have been raised about the sufficiency of global fresh water supplies and the potentially devastating impact of current and future water shortages \citep{Voeroesmarty2000, Olmstead2010, Voeroesmarty2010, Schewe2013, Vliet2021, Haefen2023}. 
Consequently, the allocation of river water is a problem of increasing importance. The aim of this paper is to study this problem. We introduce rules for a fair allocation of the rights to the water that flows along a river among the agents (countries, cities, firms) located therein. Following \cite{Ambec02} and \cite{Brink2012}, we shall consider the case of an international river, where the agents are countries located along the river, and there are tributaries on every agent's territory.\footnote{See also \cite{Ambec2008, beal2013, Brink2014, Gudmundsson2019} and \cite{Wang2022}, among others.} 
But, contrary to them, we shall not focus on the welfare agents enjoy thanks to the water, as we shall not assume agents are endowed with utility functions to transform water into (interpersonally comparable) welfare. In other words, our analysis will be \textit{resourcist} instead of \textit{welfarist}. The main reason for our modeling choice is to deal with the most fundamental and basic aspect of water allocation; namely, the allocation of riparian water rights, to be understood as the right of a property owner to use water that run through their property.

We endorse the view that fair allocation is properly modeled as a two-stage process \citep[e.g.,][]{Thomson1983, Ju2017}. 
In the first stage, property rights are fairly assigned under resource constraints. This assignment is not based on preferences, which renders the procedure informationally simple. In the second stage, \emph{voluntary exchange} from the rights assigned in the first stage is allowed. This exchange relies on individual preferences and endowments but it is not restricted by resource constraints, other than \emph{individual rationality}. 
Indeed, we believe the river sharing problem should be solved via a two-stage market-based procedure, starting with a fair allocation of riparian rights in which individual preferences should play no role.\footnote{Furthermore, although the literature typically estimates individual utilities based on water use \citep[e.g.,][]{Wu2006}, when preferences are rationalized via utilities that model the process to transform water into welfare, it would be extremely difficult to elicit them.} That first stage will be the object of our analysis.\footnote{\cite{Oeztuerk2020} takes a different approach starting from an initial (possibly unfeasible) assignment of riparian rights $r$ to obtain social welfare functions providing a complete ranking of the allocations based on the vector of $r$-equivalents (the amount of money that, when consumed with the agent’s entitlement, leaves the agent indifferent to its bundle assigned by the allocation). In that sense, her approach can be used to combine the rules we obtain in our analysis with efficiency and core constraints.} 

Our approach is thus similar to the so-called cap-and-trade systems to deal with greenhouse gas (GHG) emissions, which establish a price on emissions of (GHG-emitting) entities through a market for tradable emissions rights (called permits or allowances), achieving aggregate pollution control targets at minimum cost \citep[e.g.,][]{schmalensee2017lessons}. The assignment of permits is solved first, and the ensuing individually rational market exchange of those permits yields the final outcome.

More precisely, in our model a set of agents are located along a river (with a linear structure). For each agent, there is a river inflow. The aim is to provide rules that associate with each profile of inflows an allocation indicating the amount of water each agent uses. We impose from the outset that allocations are \emph{non-wasteful} and \emph{feasible}. That is, the river flow is fully distributed and the amount each agent gets cannot be above the overall inflow from her location upstream. 
The rules we shall consider formalize several well-known principles to deal with water disputes \citep[e.g.,][]{Kilgour1995}. 
The principles and solutions will be based solely on agents' inflows and locations.

One is the so-called principle of Absolute Territorial Sovereignty (ATS), also known as the Harmon doctrine. This principle favors upstream agents in that it allows them to use any water that flows into the river on their territory without taking into account the downstream consequences. In our setting, this will be formalized by the so-called \textit{no-transfer} rule, which simply suggests the initial profile of inflows as the solution of the allocation problem (i.e., the identity rule). 

A somewhat polar principle is the so-called principle of Unlimited Territorial Integrity (UTI). This principle favors downstream agents by stating that the inflow at the territory of some country can be claimed by all downstream countries. This can be formalized in various ways and, in our setting, we shall do so in an innovative way, via the so-called \textit{egalitarian full-transfer} rule, which allocates each inflow equally among the downstream agents (excluding the agent owning the inflow). 

A third principle is the so-called Territorial Integration of all Basin States (TIBS). 
This principle states that the water belongs to all countries together, instead of considering any country the owner of the water. That is, each country is entitled to an equal share of the river waters, regardless of its geographic location and where water enters the river. This is clearly not possible in our setting due to the feasibility constraints. We thus consider two possible interpretations for this principle in our setting. 

One is that the inflow at some country should be shared equally by this country and all its downstream countries.\footnote{Note that including the own agent in the sharing process is what constitutes the critical difference with respect to the \textit{egalitarian full-transfer} rule described above.} This gives rise to what we call the \textit{Shapley} rule, as its expression is akin to the famous value in the literature on cooperative games.

Now, the TIBS principle can also be seen as a compromise between the previous two principles (ATS and UTI). As such, we could formalize it in our setting in a different way. More precisely, we do so by means of a family of rules emerging from compromising between the \textit{no-transfer} rule and the \textit{egalitarian full-transfer} rule (via averaging the outcomes obtained by the two rules in each problem).

In our analysis, we follow the axiomatic approach, which has a long tradition in economics that can be traced back to \cite{Nash1950} and \cite{Arrow1951}. 
That is, rather than selecting among the previous rules directly, we propose to do so based on the axioms (properties) they satisfy. The selection is simplified thanks to the characterization results we provide, which are based on several axioms. 
First, the axiom of \textit{scale invariance}, stating that if all inflows are multiplied by a same positive real number, then the allocation does so too. Second, a pair of axioms modelling the effect of an increase in an agent's inflow on downstream and upstream agents. Specifically, \emph{downstream impartiality} states that the increase affects equally all equal downstream agents, whereas \emph{upstream invariance} states that upstream agents are not affected by the increase. Third, a pair of axioms focusing on the source of rivers, to be understood as their most upstream location with a positive inflow. On the one hand, \emph{balance} refers to the simplest possible scenario, in which the source is 
the only river location with non-null inflow. 
For such a scenario, the axiom states the source should receive the average amount obtained by the downstream agents. 
On the other hand, 
\emph{equal treatment of equal source inflows} states that if we have two rivers whose sources have the same inflow then these sources' assignments should be the same.

Our main results are characterizations of the two main options we considered to model the TIBS principle. 
That is, we characterize the Shapley rule and the compromise rules. It turns out that the axioms from the last pair characterize each of them, when combined with the other three axioms. More precisely, the combination of scale invariance, downstream impartiality and upstream invariance with balance characterizes the Shapley rule, whereas the combination of those first three axioms with equal treatment of equal source inflows characterizes the family of compromise rules. When we drop the two axioms referring to the sources, we also characterize a broad family of rules (obviously encompassing the previous ones). This whole family shrinks to only the no-transfer rule when we add the axiom of \textit{order preservation} to the first three axioms. This last axiom states that if an upstream agent has a higher inflow than another agent located downstream, then the assignment to the former cannot be lower than the assignment to the latter.

We extend our axiomatic analysis exploring the axiom of \emph{equal treatment of equal upstream total inflow}. This axiom states that the amount an agent gets depends on the overall available water upstream, but not on its distribution. It turns out that the combination of this axiom with the first three axioms mentioned above characterizes an alternative family of compromise rules, dubbed \textit{partial compromise} rules. More precisely, consider the \textit{egalitarian partial-transfer} rule, which represents a natural alternative to the egalitarian full-transfer rule, as it behaves as follows. Suppose all agents split their inflows in $n-1$ equal parts (where $n$ is the overall number of agents). Then, they transfer one of these parts to each of the downstream agents, but they keep the remaining parts (which would correspond to the upstream agents) for themselves. 
The egalitarian partial-transfer rule yields for each agent the aggregation of all these amounts. 
The family of \textit{partial compromise} rules is made of the convex combinations of the egalitarian partial-transfer rule and the no-transfer rule.  

We complete our study with an empirical application to the Nile River, probably one of the most famous rivers worldwide. 
In spite of the so-called Nile River Agreements that regulate water rights among countries, tensions persist, with downstream nations (such as Egypt and Sudan) demanding their share, while upstream states (such as Tanzania, and South Sudan) arguing that these agreements hinder their development. Using data from AQUASTAT, we compute the inflows of Tanzania, Uganda, South Sudan, Sudan, and Egypt, as well as the amount of water they are actually extracting from the river. Significant disparities exist between the water rights that would result from the application of the rules we introduce (and characterize) in this paper and the actual allocation of water that occurs nowadays. Our analysis implies that the actual allocation of water, emanating from the Nile River Agreements, is much closer to the UTI principle than to the ATS principle. 
Specifically, the actual allocation is closer to the outcome the egalitarian full-transfer rule would yield than to the outcome the no-transfer rule would yield. Similarly, by comparing the disparities between the families of compromise and partial compromise rules with the actual allocation, we conclude that the spirit of the Nile River Agreements is more aligned with the axiom of equal treatment of equal source inflows (which characterizes the former family) than with the axiom of equal treatment of equal upstream total inflow (which characterizes the latter family).

The rest of the paper is organized as follows. In Section 2, we introduce the model, as well as the main axioms and rules of our analysis. In Section 3, we present our main characterizations. In Section 4, we explore the alternative approach with partial transfers, which gives rise to an extra characterization. In Section 5, we present the illustration to the case of the Nile River. We conclude in Section 6. For a smooth passage, all proofs have been relegated to an Appendix. 

\section{The model}\label{model}

A set of \textbf{agents} $N=\{1, \ldots, n\}$ are located along a river, which has a linear structure. Lower numbers represent more upstream locations, so that agent 1 is the most upstream agent, agent $n$ is the most downstream agent, and $i\le j$ means that agent $i$ is upstream of agent $j$. 
For each $i \in N$, there is a river inflow, denoted by $e_i \geq 0$. Let $e=\left(e_1, \ldots, e_n\right) \in \mathbb{R}^n_+$ be the \textbf{profile of inflows}. Let $\mathcal{D}$ denote the domain of all profiles of inflows. 

Our aim is to provide rules that associate with each profile of inflows an \textbf{allocation} of riparian water rights. That is, another profile indicating the amount of water over which each agent gets the right. As water flows downstream, we require that allocations are \emph{non-wasteful} and \emph{feasible}, i.e.,  $x=\left(x_1, \ldots, x_n\right) \in \mathbb{R}^n_+$ is such that $\sum_{i=1}^n x_i=\sum_{i=1}^n e_i$ and $\sum_{i=1}^k x_i\le\sum_{i=1}^k e_i$, for each $k=1,\dots, n-1$. A \textbf{rule} $R:\mathcal{D}\to \mathbb{R}^n_+$ is a mapping that associates to each profile of inflows $e \in \mathcal{D}$ an allocation $R(e)\in \mathbb{R}^n_+$. 

The rules we shall consider formalize several well-known principles to deal with water disputes. 
One is the so-called principle of Absolute Territorial Sovereignty (ATS), also known as the Harmon doctrine. 
It states that an agent has absolute sovereignty over the area of any river basin on its territory. In our setting, this gives rise to the so-called no-transfer rule, formally defined as follows:

\textbf{No-transfer rule (Absolute Territorial Sovereignty)}. For each $e \in \mathcal{D}$, and each $i \in N$,
$$
R^{NT}_i(e) = e_i.
$$

In contrast, the so-called principle of Unlimited Territorial Integrity (UTI) favors downstream agents by conferring a right on each agent to demand the full flow of water from upstream agents. 
In our setting, this could be interpreted via the so-called egalitarian full-transfer rule, formally defined as follows:

\textbf{Egalitarian full-transfer rule (Unlimited Territorial Integrity)}. For each $e \in \mathcal{D}$, and each $i= 2,3,\dots n-1$,
$$
R^{EFT}_i(e) = \frac{e_1}{n-1}+\frac{e_2}{n-2}+\frac{e_3}{n-3}\dots \frac{e_{i-1}}{n-i+1}=\sum_{j<i} \frac{e_j}{n-j},
$$
and
$$
R^{EFT}_1(e) =0,\qquad  R^{EFT}_n(e) = \sum_{j<i} \frac{e_j}{n-j} + e_n.
$$

Both principles presented above, as well as the rules formalizing them, can be seen as rather extreme. Thus, it would be interesting to compromise among them. A plausible option is the so-called Territorial Integration of all Basin States (TIBS), which accords to each agent equal use, without regard to its contribution to the flow. The TIBS principle does not consider any country the owner of the water, but instead states that the water belongs to all countries together, no matter where it enters the river. In our setting, two possible formalizations of this principle can be considered. On the one hand, the so-called Shapley rule, formally defined as follows:\footnote{We name the rule after \cite{Shapley1953} as it coincides with the Shapley value of a suitably associated TU-game to our problem. 
The rule is also somewhat reminiscent of the so-called serial cost-sharing rule introduced by \cite{Moulin1992}.}

\textbf{Shapley rule (Territorial Integration of all Basin States)}. For each $e \in \mathcal{D}$ and each $i\in N$,
$$ 
R^{Sh}_i(e) = \frac{e_1}{n}+\frac{e_2}{n-1}+\frac{e_3}{n-2}+\ldots+\frac{e_i}{n-i+1} = \sum_{j \leq i} \frac{e_j}{n-j+1}. 
$$

Note that this rule is similar to the previous one, in the sense that it imposes agents equal transfers downstream. The crucial difference between both is that with this rule each agent also keeps an equal share before transferring water downstream.

On the other hand, the rules emerging from averaging between the two rules introduced above (which were
formalizing the ATS and UTI principles, 
respectively). Formally,

\textbf{Compromise rules (Territorial Integration of all Basin States)}. For each $\lambda\in[0,1]$ and each $e \in \mathcal{D}$, 
$$
R^{\lambda}(e) = \lambda R^{NT}(e) + (1-\lambda)R^{EFT}(e).
$$

We illustrate all these rules in the following example.
\begin{example}
Consider the profile of inflows $e=(50,30,10,10)$. The no-transfer rule obviously yields the allocation $(50,30,10,10)$. The Shapley rule yields $1/4$ of the first inflow $(50)$ to each agent; $1/3$ of the second inflow $(30)$ to each of the last three agents; $1/2$ of the third inflow $(10)$ to each of the last two agents; and the last agent keeps her own inflow $(10)$. The egalitarian full-transfer rule yields $1/3$ of the first inflow $(50)$ to each of the last three agents; $1/2$ of the second inflow $(30)$ to each of the last two agents; the third inflow $(10)$ to the last agent; and the last agent keeps her own inflow $(10)$ too. Finally, the compromise rule yields the convex combination between the allocations provided by the no-transfer rule and the egalitarian full-transfer rule. 
    \begin{center}
    \begin{tabular}{ccccc}
    \toprule
    & \multicolumn{4}{c}{Allocations} \\ \cmidrule{2-5}
    Rule & $x_1$ & $x_2$ & $x_3$ & $x_4$ \\ 
    \midrule
    $R^{NT}$ & $50$ & $30$ & $10$ & $10$\\[0.3cm]
    $R^{Sh}$ & $\frac{50}{4}$ & $\frac{50}{4}+10$ & $\frac{50}{4}+10+5$ & $\frac{50}{4}+10+5+10$\\[0.3cm]
    $R^{EFT}$ & $0$ & $\frac{50}{3}$& $\frac{50}{3}+15$ & $\frac{50}{3}+15+10+10$\\[0.3cm]
    $R^{\lambda}$ & $50\lambda$ & $\frac{50}{3}+\frac{40}{3}\lambda$& $\frac{95}{3}-\frac{65}{3}\lambda$ & $\frac{155}{3}-\frac{125}{3}\lambda$\\
    \bottomrule
    \end{tabular}
    \end{center}
\end{example}
\bigskip

Instead of endorsing one rule or another directly, we follow the axiomatic approach to derive rules. That is, we formalize as axioms principles that are normatively appealing (either from an ethical, or an operational perspective). We then study the implications of those axioms, hoping to characterize specific rules by combining some of the axioms. 

We start our inventory of axioms with a standard axiom. If all inflows are multiplied by a same positive real number, then the allocation is too. 

\textbf{Scale invariance}. For each $e \in \mathcal{D}$ and each $\gamma \in \mathbb{R}_+$, $R(\gamma e)=\gamma R(e)$.

Our next two axioms refer to situations in which the inflow of an agent increases. How should this affect the other agents? The answer will depend on whether we focus on upstream agents or downstream agents (from the location in which the increase occurs). \emph{Upstream invariance} states that this change does not alter the allocation of upstream agents. Formally, 

\textbf{Upstream invariance}. For each pair $e,e' \in \mathcal{D}$, such that $e_i<e'_i$ for some $i \in N$, and $e_j = e'_j$ for all $j \in N \backslash \{i\}$, then 
for each $k<i$,
$$
R_k(e) = R_k(e').
$$

For agents downstream, our axiom only pertains to agents who have equal inflows.\footnote{The axiom could be strengthened to all downstream agents (even without different inflows). Our results would also hold in that case.} 
Specifically, \emph{downstream impartiality} states that, if two downstream agents have equal inflows, then they are equally affected.

\textbf{Downstream impartiality}. For each pair $e,e' \in \mathcal{D}$, such that $e_i<e'_i$ for some $i \in N$, $e'_j = e_j$ for all $j \in N \backslash \{i\}$, and $e_k=e_l$ for all $k,l>i$, then
$$
R_k(e') - R_k(e) = R_l(e') - R_l(e).
$$

The next axiom, \textit{order preservation}, is also a standard axiom stating that if an agent has a higher inflow than another agent downstream, then the amount the former obtains cannot be lower than the amount the latter obtains.

\textbf{Order preservation}. For each $e \in \mathcal{D}$, if $i<j$ and $e_i \geq e_j$ then $R_i(e) \geq R_j(e)$.

To conclude, we introduce axioms referring to the source of a river, to be understood as its most upstream location with a positive inflow. More precisely, for each $e \in \mathcal{D}$, its \textit{source} is $s(e)=\min\{k\in \{1,2,\dots n-1\}: e_k>0\}$. 
Note that all upstream agents of a source have zero inflows. Thus, they cannot receive water from the source. But the source agent can obviously transfer water downstream, even when all downstream agents also have zero inflows. 
In such a case, it seems pertinent to compare the amount obtained by the source agent 
and the average amount obtained by the downstream agents. If the former is lower, we shall say the rule is \emph{progressive}. If the latter is lower, we shall say the rule is \emph{regressive}. If both coincide, we shall say that the rule is \emph{balanced}. Formally, 

\textbf{Progressivity}. For each $e \in \mathcal{D}$, such that $e_i>0$ for some $i \in \{1,\dots, n-1\}$, and $e_j=0$ for all $j \in N \backslash \{i\}$, then
$$
R_i(e)\le\frac{1}{n-i}\sum_{k>i} R_k(e).
$$

\textbf{Regressivity}. For each $e \in \mathcal{D}$, such that $e_i>0$ for some $i \in \{1,\dots, n-1\}$, and $e_j=0$ for all $j \in N \backslash \{i\}$, then
$$
R_i(e)\ge\frac{1}{n-i}\sum_{k>i} R_k(e).
$$

\textbf{Balance}. For each $e \in \mathcal{D}$, such that $e_i>0$ for some $i \in \{1,\dots, n-1\}$, and $e_j=0$ for all $j \in N \backslash \{i\}$, then
$$
R_i(e)=\frac{1}{n-i}\sum_{k>i} R_k(e).
$$

Finally, we consider the axiom stating that if we have two rivers whose sources have the same inflow then these sources receive the same amounts.  

\textbf{Equal treatment of equal source inflows}. For each pair $e,e' \in \mathcal{D}$, such that $e_{s(e)}=e'_{s(e')}$,
$$
R_{s(e)}(e)  = R_{s(e')}(e').
$$

\section{The main results}

We provide our main results in this section. First, a characterization of the Shapley rule, one of our proposals to model the TIBS principle. 

\begin{theorem}\label{ShTh}
    A rule satisfies scale invariance, downstream impartiality, upstream invariance, and balance if and only if it is the Shapley rule.
\end{theorem}

If we replace balance in the previous result by equal treatment of equal source inflows, we characterize the compromise rules, our alternative proposal to model the TIBS principle. 
 
\begin{theorem}\label{FamilyTh}
    A rule satisfies scale invariance, downstream impartiality, upstream invariance, and equal treatment of equal source inflows if and only it is a compromise rule. 
\end{theorem}

If we consider instead order preservation, we characterize the rule modeling the ATS principle, 
as stated in the next result. 

\begin{theorem}\label{NTTh}
A rule satisfies scale invariance, downstream impartiality, upstream invariance, and order preservation if and only it is the no-transfer rule.
\end{theorem}

Our most general result characterizes all the rules that satisfy the three common axioms in the previous results; namely, scale invariance, downstream impartiality and upstream invariance. This is a large family of rules (which obviously includes both the Shapley rule and the compromise rules characterized above) depending on $n-1$ parameters: $\alpha=(\alpha_1, \ldots, \alpha_{n-1}) \in [0,1]^{n-1}$. The rule $R^{\alpha}$ works as follows. Agent $1$ keeps a portion $\alpha_1$ of the first inflow ($e_1$) and transfers the rest ($(1-\alpha_1)e_1$) equally among the $n-1$ remaining agents downstream. Similarly, agent $2$ keeps a portion $\alpha_2$ of the second inflow ($e_2$) and transfers the rest ($(1-\alpha_2)e_2$) equally among the $n-2$ remaining agents downstream. And so on, until the last agent ($n$) who keeps the whole last inflow ($e_n$).\footnote{This is somewhat reminiscent to the so-called $\alpha$-procedure, recently introduced by \cite{Brandenburger2023}.} Note that if the parameters are all equal to $0$, we obtain the egalitarian full-transfer rule; if they all equal to $1$, we obtain the no-transfer rule; and if they are all equal to $\lambda\in(0,1)$ we obtain the compromise rule $R^{\lambda}$. Finally, if for each $k=1,\dots n-1$, $\alpha_k=\frac{1}{n-k+1}$, we obtain the Shapley rule.

\begin{theorem}\label{thm1}
A rule $R$ satisfies scale invariance, downstream impartiality, and upstream invariance if and only if it is an $R^{\alpha}$ rule. That is, there exist $\alpha=(\alpha_1, \ldots, \alpha_{n-1}) \in [0,1]^{n-1}$ such that
\begin{align*}
R(e) =& \left( \alpha_1 e_1, \alpha_2 e_2+\frac{e_1\left(1-\alpha_1\right)}{n-1}, \alpha_3 e_3+\frac{e_1\left(1-\alpha_1\right)}{n-1}+\frac{e_2\left(1-\alpha_2\right)}{n-2}, \ldots, \right.\\
    & \left. e_n+\frac{e_1(1-\alpha)}{n-1}+\ldots+\frac{e_{n-1}\left(1-\alpha_{n-1}\right)}{n-(n-1)} \right)\equiv R^{\alpha}(e).
\end{align*}

\end{theorem}

As stated in Theorem \ref{ShTh}, if we add balance to the axioms of Theorem \ref{thm1}, then we characterize the Shapley rule. If, instead, we add progressivity, then we characterize the sub-family of rules $\{R^{\alpha}: \alpha_k\leq\frac{1}{n-k+1}\text{ for each }k=1,\dots,n-1\}$. Likewise, if we add regressivity, then we characterize the sub-family of rules $\{R^{\alpha}: \alpha_k\geq\frac{1}{n-k+1}\text{ for each }k=1,\dots,n-1\}$. 

Finally, as stated in Theorem \ref{FamilyTh}, if we add equal treatment of equal source inflows to the axioms of Theorem \ref{thm1}, then we characterize the compromise rules $\{R^{\lambda}: \lambda \in [0,1]\}$. If we also add progressivity, then we characterize the sub-family of compromise rules $\{R^{\lambda}: \lambda \in [0,\frac{1}{n}]\}$. Likewise, if we add regressivity, then we characterize the sub-family of compromise rules $\{R^{\lambda}: \lambda \in [\frac{1}{2},1]\}$. 


\section{Further results}\label{further}

In this section, we extend our analysis in the previous section to account for an alternative approach to the UTI and TIBS principles. 
More precisely, we start from the premise that, in each problem, all agents split their inflows in $n-1$ equal parts. Then, they transfer one part to each of the downstream agents, but keep the remaining parts (which would correspond to the upstream agents) for themselves. The following rule yields as an allocation for the problem the aggregation of all those amounts. 

\textbf{Egalitarian partial-transfer rule}. 
For each $e \in \mathcal{D}$, and each $i \in N$. 
$$
R^{EPT}_i(e) = \left[1-\frac{(n-i)}{n-1}\right] e_i + \frac{1}{n-1} \sum_{k<i} e_k.
$$

We can also define a family of \textit{partial compromise rules} by means of convex combinations between the rule just defined and the no-transfer rule.\footnote{Note that the unique intersection between both families of compromise rules is the no-transfer rule.} Formally, 

\textbf{Partial compromise rules}. For each $\delta\in[0,1]$, and each $e \in \mathcal{D}$, 
$$
R^{\delta}(e) = \delta R^{NT}(e) + (1-\delta)R^{EPT}(e).
$$

If we now revisit Example 1 from the previous section, and consider the profile of inflows $e=(50,30,10,10)$, we obtain that the egalitarian partial-transfer rule yields $1/3$ of the first inflow ($50$) to each of the last three agents; $1/3$ of the second inflow ($30$) to each of the last two agents, while keeping $1/3$ of it for the second agent; $1/3$ of the third inflow ($10$) to the last agent, while keeping $2/3$ of it for the third agent; and the last agent keeps her own inflow ($10$). The partial compromise rule yields the convex combination between the allocations provided by the no-transfer rule, $(50,30,10,10)$, and the egalitarian partial-transfer rule.  
\medskip

\begin{center}
    \begin{tabular}{ccccc}
    \toprule
    & \multicolumn{4}{c}{Allocations} \\ \cmidrule{2-5}
    Rule & $x_1$ & $x_2$ & $x_3$ & $x_4$ \\ 
    \midrule
    $R^{NT}$ & $50$ & $30$ & $10$ & $10$\\[0.3cm]
    $R^{EPT}$ & $0$ & $\frac{50}{3}+10$& $\frac{50}{3}+10+\frac{20}{3}$ & $\frac{50}{3}+10+\frac{10}{3}+10$\\[0.3cm]
    $R^{\delta}$ & $50\delta$ & $\frac{80}{3}+\frac{10}{3}\delta$& $\frac{100}{3}-\frac{70}{3}\delta$ & $40-30\delta$\\
    \bottomrule
    \end{tabular}
    \end{center}
\bigskip

Now, to illustrate the connection of both families, we consider the following example. 
\begin{example}
    Let $e=(3,2,1,1)$, $\lambda\in[0,1]$ and $R^\lambda$ be the corresponding compromise rule. Such a rule guarantees each agent the $\lambda$ share of its own inflow, and transfers the rest downstream to be equally split among the (downstream) agents. In particular,
    \begin{itemize}
        \item Agent 1 keeps $3\lambda$ of her own inflow and transfers $3(1-\lambda)$ downstream. Thus, $x_1=3\lambda$.
        \item Agent 2 keeps $2\lambda$ of her own inflow and transfers $2(1-\lambda)$ downstream. In addition, agent 2 receives $\frac{3(1-\lambda)}{3}$ from agent 1. Thus, $x_2=2\lambda + \frac{3(1-\lambda)}{3}=1+\lambda$.
        \item Agent 3 keeps $\lambda$ of her own inflow and transfers $(1-\lambda)$ downstream. In addition, agent 3 receives $\frac{3(1-\lambda)}{3}$ from agent 1 and $\frac{2(1-\lambda)}{2}$ from agent 2. Thus, $x_3=\lambda +\frac{3(1-\lambda)}{3} + \frac{2(1-\lambda)}{2}=2-\lambda$.
        \item Finally, agent 4 keeps her whole inflow because she is the last agent. In addition, agent 4 receives $\frac{3(1-\lambda)}{3}$ from agent 1, $\frac{2(1-\lambda)}{2}$ from agent 2, and $(1-\lambda)$ from agent 3. Thus, $x_4=1 + \frac{3(1-\lambda)}{3} + \frac{2(1-\lambda)}{2} + (1-\lambda)=4-3\lambda$.
    \end{itemize}

    Let $\beta\in[0,1]$ and consider the corresponding partial compromise rule $R^\beta$. Such a rule guarantees each agent a fixed share $\beta$ of the inflows of each of her predecessors, rather than a fixed share of her own inflow. In particular,
    \begin{itemize}
        \item Agent 4 keeps her own inflow (she is the last agent) and receives $\frac{1}{3}\beta $ from agent 3, $\frac{2}{3}\beta $ from agent 2, and $\frac{3}{3}\beta $ from agent 1. Thus, $x_4=1 + \left(\frac{1}{3}\beta  + \frac{2}{3} \beta +\frac{3}{3}\beta  \right) = 1+2\beta$.
        \item Agent 3 receives $\frac{2}{3}\beta $ from agent 2, and $\frac{3}{3}\beta $ from agent 1. Regarding her own inflow, in the previous step she has already transferred $\frac{1}{3}\beta $ to agent 4; therefore, she keeps for herself $1-\frac{1}{3}\beta$. Thus, $x_3=1- \frac{1}{3}\beta +\left( \frac{2}{3} \beta+ \frac{3}{3}\beta \right) = 1 + \frac{4}{3}\beta$.
        \item Agent 2 receives $\frac{3}{3}\beta$ from agent 1. Regarding her own inflow, in the previous steps she has already transferred $\frac{2}{3}\beta$ to agent 4 and $\frac{2}{3}\beta$ to agent 3, therefore, she keeps for herself $2-2 \frac{2}{3}\beta$. Thus, $x_2=2-2 \frac{2}{3}\beta +\frac{3}{3} \beta = 2 - \frac{1}{3}\beta$.
        \item Finally, agent 1 does not receives anything from her predecessors because she in the first agent in the river. Regarding her own inflow, in the previous steps she has already transferred $ \frac{3}{3}\beta$ to agent 4, $\frac{3}{3}\beta$ to agent 3 and $\frac{3}{3}\beta$ to agent 4, therefore, she keeps for herself $3-3\frac{3}{3}\beta=3$. Thus, $x_1=3-3\beta$.
    \end{itemize}
\end{example}

With any partial compromise rule, the allocation of an agent depends on the overall available water upstream, but not on its distribution. That is, those rules satisfy the following axiom. 

\textbf{Equal treatment of equal upstream total inflow}. For each pair $e,e' \in \mathcal{D}$ and each $i \in N$, if $e_i=e'_i$ and $\sum_{j<i} e_j = \sum_{j<i} e'_j$ then 
$$
R_i(e)=R_i(e').
$$

Note that the above axiom is an alternative to that of equal treatment of equal source inflows in the previous section. As stated in the next result, if we replace the latter by the former in Theorem \ref{FamilyTh}, we obtain a new characterization. In other words, this new axiom characterizes the family of partial compromise rules, when combined with our three structural axioms.

\begin{theorem}\label{thm6}
A rule $R$ satisfies scale invariance, downstream impartiality, upstream invariance, and equal treatment of equal upstream total inflow if and only if it is a partial compromise rule. 
\end{theorem}

\section{A case study: the Nile River}

The Nile River is considered the largest river in the world, with a length of about 6,650 kilometers. It rises south of the Equator and flows northward through northeastern Africa to drain into the Mediterranean Sea. Its basin includes parts of Tanzania, Burundi, Rwanda, the Democratic Republic of the Congo, Kenya, Uganda, South Sudan, Ethiopia, Sudan, and the cultivated part of Egypt. About 250 million people rely on the river for daily water, and approximately 10\% of them already face water scarcity. The UN projects that this percentage will increase to 35\% (around 80 million people) in 2040. It is not surprising that the distribution of the Nile’s water supply has been a constant source of conflicts. A series of colonial-era agreements affect use of the Nile River. Two commonly cited agreements in terms of water allocation and the purported rights of riparians include a 1929 Exchange of Notes between His Majesty’s Government in the United Kingdom and the Egyptian Government in Regard to the Use of the Waters of the River Nile for Irrigation Purposes, and the 1959 Agreement between the Republic of Sudan and the United Arab Republic (of Egypt) for the Full Utilization of the Nile Waters.\footnote{\url{http://www.internationalwatersgovernance.com/nile-river-basin-initiative.html}. Last accessed July 18th, 2024.}
These agreements dictate that upstream riparian states (such as Kenya, Tanzania and South Sudan) must respect the water rights of downstream countries (primarily Egypt and Sudan). Additionally, they are prohibited from constructing dams or initiating river projects without approval from downstream nations, particularly Egypt. The upstream countries have consistently disputed and questioned these treaties. While Egypt and Sudan demand that their water share must be upheld, the upstream states argue that these agreements are unfair and hinder harm their agricultural and developmental objectives.\footnote{\url{https://arabcenterdc.org/resource/water-conflict-between-egypt-and-ethiopia-a-defining-moment-for-both-countries/}. Last accessed July 18th, 2024.} 
The so-called Nile Basin Initiative was launched in February 1999 by the water ministers of the countries that share the river with the aim of “seeking to develop the river in a cooperative manner, share substantial socioeconomic benefits, and promote regional peace and security” and to “provide[] an institutional mechanism, a shared vision, and a set of agreed policy guidelines to provide a basinwide framework for cooperative action.”

In this section, we resort to the rules we introduced (and characterized) above in order to illustrate how they could be applied to help achieving the goals of the Nile Basin Initiative. We acknowledge that the issue of allocating riparian water rights in the Nile River is more complex, given the historical, political, cultural, economic, as well as social ramifications. We nevertheless believe that our empirical illustration, limited as might be by the theoretical model we consider (thus not encompassing all relevant factors at stake), 
still offers valuable insights.

\begin{figure}[h]
  \centering
  \includegraphics[scale=0.7]{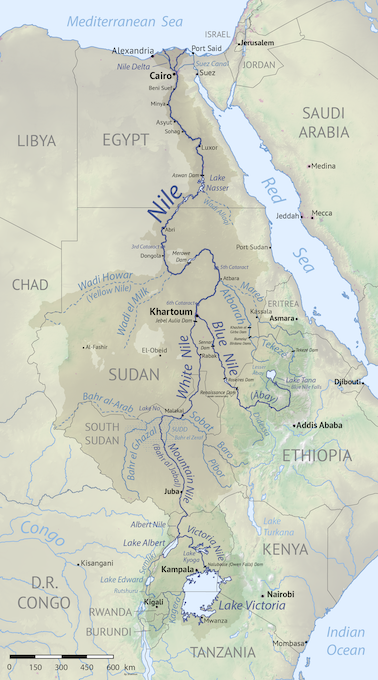}
  \caption{Nile River Basin \label{fig:NileBasin}}
\end{figure}

To be consistent with the theoretical model presented in the previous section, we consider the part of the Nile River that flows through Tanzania, Uganda, South Sudan, Sudan and Egypt, in that particular order; thus being Tanzania and Egypt the most upstream and downstream countries, respectively (see Figure \ref{fig:NileBasin}).\footnote{To include all the nations in the Nile Basin, we would need to expand the model to account for agents located in more complex structures than a simple line.} That is, in the parlance of our model, $N=\{1,2,3,4,5\}=\{ \text{Tanzania}, \text{Uganda}, \text{South Sudan}, \text{Sudan}, \text{Egypt}\}$. 

We resort to AQUASTAT to compute the inflows and actual allocations of the six countries we study.\footnote{AQUASTAT, developed by the Food and Agriculture Organization (FAO), is a global information system focused on water resources and agricultural water management. It compiles, analyzes, and provides free access to over 180 variables and indicators by country. Free access to data is provided at \url{https://www.fao.org/aquastat/en/databases/}} 

As for inflows, Lake Victoria contributes a substantial 33 km$^3$/year to the Nile. Its surface is divided among three countries: Tanzania (51\%), Uganda (43\%) and Kenya (6\%). Tanzania's inflow accounts for 51\% of Lake Victoria's contribution to the Nile. Uganda's inflow is a combination of its portion of Lake Victoria's contribution (based on surface area) and water from Lake Albert, estimated at 2 km$^3$/year. The two main tributaries of the Nile River in South Sudan are the Bahr al Qhazal and Sobat Rivers. The first one flows almost entirely within South Sudan and discharges 1.5 km$^3$/year at its mouth. The second tributary is the confluence of the Pibor and Baro Rivers, whose contributions are 3.1 and 13 km$^3$/year, respectively. Thus, the overall South Sudan's inflow is 17.6 km$^3$/year. The forth country in the course, Sudan, receives inputs from the Blue Nile (flowing from Ethiopia) and the Albarath River (also originating in Ethiopia and Eritrea). Together, these contribute 65.3 km$^3$/year to the Nile. Finally, according to AQUASTAT, Egypt completely relies on the inflows from upstream countries, as it does not provide water to the river. The resulting amounts described in this paragraph for the five countries are listed in the second column of Table \ref{table2}. 

To estimate the actual water allocation among the countries along the Nile River, we have considered as a proxy the so-called \emph{total freshwater withdrawal} variable in AQUASTAT. The withdrawals from Tanzania, Uganda, South Sudan, Sudan, and Egypt amount to 5.18 km$^3$/year, 0.64 km$^3$/year, 0.66 km$^3$/year, 26.93 km$^3$/year and 77.7 km$^3$/year, respectively. The cumulative allocation of these withdrawals totals 110.9km$^3$/year, which is slightly below the overall inflow of 115.9 km$^3$/year. To be consistent with our theoretical model, we have proportionally adjusted the withdrawals so that the total sum is also 115.9 km$^3$/year. These adjustments, which only result in minor differences, are listed into the third column of Table \ref{table2}.

The allocations of riparian water rights of the Nile River according to the rules we introduced are gathered in the remaining columns of Table \ref{table2}. We consider three rules from each of the two families we characterized (in Theorems \ref{FamilyTh} and \ref{thm6}), as well as the Shapley rule (which we also characterized in Theorem \ref{ShTh}). A first obvious aspect to notice is that Tanzania, Uganda, and South Sudan exhibit similar inflows. However, the allocation for these countries varies significantly across rules (except, obviously, for the no-transfer rule).
Egypt, in spite of having the highest actual allocation, receives nothing under the no-transfer rule (sixth column) due to its lack of contribution to the river. Conversely, the egalitarian full-transfer (fourth column) rule greatly benefits Egypt. The Shapley rule (last column) assigns equal amounts of water to both Sudan and Egypt, which are the two most downstream locations along the river, respectively. Sudan, being the largest contributor, shares its inflow equally with Egypt under this rule, resulting in a substantial assignment for the latter. Among the rules we consider, the egalitarian partial-transfer rule (previous to last column) is the one that best reflects Sudan's significant contribution to the Nile River, while simultaneously meeting Egypt's demand. In contrast, Tanzania receives zero under this rule. 
Finally, the middle compromise rule ($\lambda=\frac{1}{2}$) provides the most equitable water allocation (fifth column). Its counterpart (partial compromise) rule ($\delta=\frac{1}{2}$) allocates the largest flow share to Sudan, reserves water for Egypt, and still provides flow to Tanzania (seventh column).

\begin{table}[h]
\centering
\begin{tabular}{lrrrrrrrr}
\toprule
& & & \multicolumn{6}{c}{Water rights} \\
\cmidrule(lr){4-9}
Countries & $e$ & $z$ & $R^{EFT}$ & $R^{\lambda=\frac{1}{2}}$ & $R^{NT}$ & $R^{\delta=\frac{1}{2}}$ & $R^{EPT}$ & $R^{Sh}$  \\ 
\midrule 
Tanzania    & 16.8 & 5.4  & 0    & 8.40  & 16.8 & 8.40  & 0     & 3.36  \\
Uganda      & 16.2 & 0.7  & 4.2  & 10.20 & 16.2 & 12.22 & 8.25  & 7.41  \\
South Sudan & 17.6 & 0.7  & 9.6  & 13.60 & 17.6 & 17.33 & 17.05 & 13.28 \\
Sudan       & 65.3 & 28.1 & 18.4 & 41.85 & 65.3 & 63.46 & 61.62 & 45.93 \\
Egypt       & 0    & 81   & 83.7 & 41.85 & 0    & 14.49 & 28.98 & 45.93 \\
\bottomrule
\end{tabular}
\caption{Inflows, actual allocations and water-use rights in the Nile River.\label{table2}}
\end{table}

We now move to percentages of the aggregate inflow that each country enjoys under the different rules. 
For instance, Tanzania represents 14.5\% of the total inflow, and all rules (except the no-transfer rule) allocate it a smaller share. However, in all cases (except for the Shapley rule), Tanzania receives more than its actual allocation. Uganda and South Sudan contribute around 14\% and 15\%, respectively, but they receive an insignificant share of just 0.6\%. The rules offer them similar shares (except for the egalitarian partial-transfer rule), which are much larger than those from the actual allocation, but smaller than those of their inflows. As the largest contributor, Sudan supplies over half of the total Nile flow. This share is guaranteed with the three rules from the family of partial compromise rules. In addition, compared to its actual allocation, Sudan receives larger shares of water under almost any rule. The exception is the egalitarian full-transfer rule, as Sudan's inflow is equally split with Egypt. This last country is a revealing case. It contributes nothing but it currently receives about 70\% of the Nile's total resources. The egalitarian full-transfer rule further increases Egypt's share, due to its strategic location at the river's mouth, downstream from Sudan. However, other rules (not only the obvious case of the no-transfer rule) reduce the share, which is especially the case for the partial compromise rules.

Theorem \ref{FamilyTh} states that, when we require scale invariance, downstream impartiality, upstream invariance, and equal treatment of equal source inflow, then we are compelled to select from the family of compromise rules. These are convex combinations of the no-transfer and the egalitarian full-transfer rules, via the parameter $\lambda \in [0,1]$. As $\lambda$ decreases, the distribution of flow aligns more closely with the UTI principle. Conversely, higher values of $\lambda$ indicate a stronger alignment with the ATS principle. It is then natural to wonder which of these two principles governs the actual allocation. To do that, assuming the four axioms mentioned above are compelling, we compute the value of $\lambda \in [0,1]$ that minimizes the Euclidean distance between $z$ and $R^{\lambda}(e)$. This turns out to be $\lambda^*=0.068$.\footnote{$R^{\lambda^*}$ is not only the rule providing the closest allocation to $z$ within the family of compromise rules, but it is also the rule providing the closest allocation among all rules in Table \ref{table2}. Within the family of partial compromise rules, the one providing the closest allocation is the one corresponding to $\delta^*=0$, i.e., the egalitarian partial-transfer rule.} Thus, we can state that the actual allocation of water along the Nile is much closer to agree with the UTI principle than to the ATS principle. Specifically, the allocation of $R^{\lambda^*}(e)$ is:
$$
R^{\lambda^*}(e) = (1.1, 5, 10.2, 21.6, 78).
$$
Interestingly, Theorem \ref{FamilyTh} also provides a criterion to determine whether the actual allocation of water for each country is reasonable or not. Assuming the axioms characterizing the compromise rules are compelling, the two extremes of this family yield natural bounds within which a reasonable allocation of water should fall. In other words, the actual allocation of a country could be considered legitimate if it falls between the amounts determined by the no-transfer and the egalitarian full-transfer rules. On this basis, Tanzania, Sudan and Egypt receive legitimate amounts of water. On the other side, Uganda and South Sudan do not as they receive amounts that are significantly below their corresponding lower bounds. If instead of equal treatment of equal source inflow, we would find the axiom of equal treatment of equal upstream total inflow more compelling, we could perform the same analysis with the bounds the partial compromise rules yield. That is, the no-transfer and the egalitarian partial-transfer rules would now yield the limits for a reasonable allocation of water. If so, Tanzania would be the unique country receiving a legitimate amount. However, Uganda, South Sudan, and Sudan would receive less than what they should, while Egypt would receive more than it should.

To conclude with this illustration, we shift the focus from specific rules to families in order to explore their descriptive power. More precisely, we explore which family reflects the real distribution of water more accurately. To do so, we proceed as follows. For each pair $(\lambda, \delta) \in [0,1]^2$, we consider the Euclidean distances between $R^\lambda$ (Theorem \ref{FamilyTh}) and $R^\delta$ (Theorem \ref{thm6}), and the actual allocation $z$. That is, 
$$
d^{F}(\lambda)=\| R^\lambda (e)-z\|_2, 
$$
and
$$
d^{P}(\delta)=\| R^\delta (e)-z\|_2. 
$$
We then aggregate over the interval $[0,1]$. That is,
$$
\int_0^1 d^{F}(\lambda) d\lambda=46.52,
$$ 
and 
$$
\int_0^1 d^{P}(\delta) d\delta=78.27.
$$
Based on these results, we conclude that the descriptive power of the compromise family is stronger than that of the partial compromise family.\footnote{This would also be obtained with other criteria that could also be applied for distances (e.g., minmin, minmax, etc.).}
Notice that Theorems \ref{FamilyTh} and \ref{thm6} differ only in one axiom (equal treatment of equal source inflows versus equal treatment of equal upstream total inflow). Therefore, as we obtain that the actual allocation is closer to the compromise family than to the partial compromise family, then one may argue that the descriptive appeal of the axiom of equal treatment of equal source inflows prevails over the descriptive appeal of the axiom of equal treatment of equal upstream total inflow.

\section{Discussion}
We have studied in this paper the fair allocation of riparian water rights. In our stylized model, agents are located along a river and each agent is endowed with an amount of (inflow) water. The issue is to reallocate the overall inflow among them, subject to feasibility constraints and taking into account their location in the river and their individual inflows. Our results provide normative foundations for various forms of the principle of Territorial Integration of all Basin States, a reasonable compromise between the more extreme principles of Absolute Territorial Sovereignty and Unlimited Territorial Integrity. We thus highlight the role of Territorial Integration of all Basin States as a basic principle for the fair allocation or riparian water rights. 

Our model does not accommodate the possibility that agents are unequal in other relevant respects (beyond inflow and location along the river). For instance, in terms of population, which might be a relevant aspect to explore the fair allocation of riparian water rights. It is left for further research to extend our analysis in that direction. Note that some of the axioms we considered in this work would still apply in that enriched setting. But others would need to be modified to properly model the corresponding normative principles in the enriched setting.

Our analysis endorses the standard approach in the so-called bankruptcy problems \citep{ONeill1982,Thomson2019} to address the central question in economics, pertaining to the allocation of scarce and valuable resources.\footnote{See \cite{EstevezFernandez21} for an analysis of sequential bankruptcy problems, which are more similar to the model we analyze here.} 
Preferences do not reach satiation in these contexts but there are natural upper bounds on consumptions (in the case of bankruptcy problems, no agent should be assigned more than her claim and in our case of river sharing, no agent should be assigned more than her upstream inflow). 
Agents differ in our model only to the extent that their locations and inflows differ and we do not take into account the intensity of the satisfaction they derive from their assignments, as captured by 
 utility functions (which might be private information, and thus difficult to elicit). 
Utility functions, which 
convert water into welfare, would nevertheless be crucial to explain end-state allocations derived from a market-based protocol from the allocation of (riparian) rights.\footnote{Note that if we would, simplistically, assume that all agents have the same utility function then the rule in \cite{Ambec02} would amount to a constrained egalitarian rule in which riparian water rights are as equal as possible.} We do not explore that second (market-based) stage in our paper and focus on the (important) first stage of the (fair) allocation of riparian water rights.\footnote{See \cite{Weber2001}, \cite{Ansink2009}, \cite{Wang2011}, \cite{Ansink2012}, \cite{Ambec2013}, \cite{Aghakouchak2014}, \cite{Ju2017}, \cite{ferguson2024market} or \cite{Donna2024} for various aspects of the interplay between water rights management and water markets.}
Our model is thus similar to the problem of cleaning a polluted river, in which the input is the cost associated to each agent located along the river \citep[e.g.,][]{Ni2007, Dong2012, Brink2018, AlcaldeUnzu2015, AlcaldeUnzu2020, Li2022}. 
The (mathematical) difference with our model is that, beyond non-wastefulness, we also impose additional feasibility constraints for the allocations our rules yield. 
Thus, the rules used in that setting might not be well defined in ours. On the other hand, the rules we consider here would be well defined in that setting and could potentially be used therein. For instance, this is obviously the case of the no-transfer rule, which is dubbed the \textit{local responsibility sharing method} by \cite{Ni2007} in their setting. But also the case of the Shapley rule, dubbed the \textit{downstream equal sharing method} by \cite{Dong2012} in their setting. The so-called \textit{upstream equal sharing method} \citep[e.g.,][]{Ni2007}, which allocates costs in the ``opposite way" and plays a focal role in that setting, does not have a counterpart in ours, as it may well violate the feasibility constraints. 

Our model is also similar to the problem of revenue sharing in hierarchies \citep[e.g.,][]{Hougaard17}, in which the issue is to distribute the proceeds generated by agents located along a hierarchy. In the benchmark case in which the hierarchy is a line, agents are thus just characterized by the location in the line and the revenue they bring to the hierarchy. Transfer rules determine a redistribution of the overall revenue along the line, with the only caveat, beyond non-wastefulness, that revenues can only be transferred upwards in the hierarchy. As such, the model can simply be interpreted as the mirror image of ours (in which the bottom agent of the river plays the role of the agent at the top of the hierarchy) and our feasibility constraints for the allocations our rules are satisfied by the definition of (upward) transfer rules in the hierarchy. \cite{Hougaard17} characterize a family of \textit{geometric} transfer rules where revenue `bubbles up' in the hierarchy. This family ranges from the \textit{no-transfer} rule to the \textit{full-transfer} rule. In the former, no revenue bubbles up, and is thus equivalent to the namesake rule in our model. In the latter, all the revenues bubble up to the top of the hierarchy. This could be interpreted in our model as transferring the overall inflow of the river to the most downstream agent. Such a rule would be an alternative way of translating the UTI principle to the egalitarian full transfer rule, but it would violate the axiom of downstream egalitarian invariance. 
It is nevertheless left for further research to explore the counterpart of the family of geometric rules for the fair allocation of riparian water rights. 

To conclude, we mention that our model is also similar to the model recently introduced by \cite{Gudmundsson2024} to analyze the problem of sharing sequentially triggered losses. In their case, a group of agents are aligned representing a loss chain in which agent $1$ initiates the chain and 
agent $n$ is the final agent. A problem is a vector $\ell=(\ell_1,\dots\ell_n)$ in which agent $i$ causes loss $\ell_i \ge 0$ to agent $i+1$. An allocation is a list $x=\left(x_1, \ldots, x_n\right) \in \mathbb{R}^n_+$ which specifies each agent $i$’s
liability $x_i \ge 0$ and  is such that $\sum_{i=1}^n x_i=\sum_{i=1}^n \ell_i\equiv L$ and $\sum_{i=1}^k x_i\ge\sum_{i=1}^k \ell_i$, for each $k=1,\dots, n-1$. That is, exactly the opposite to our feasibility constraints as, in their case, agents later in the chain bear no responsibility for losses caused by earlier agents, whereas, in our case, water flows downstream along the river. The no-transfer rule considered here is dubbed the \textit{direct liability rule} by \cite{Gudmundsson2024} in their setting. The so-called \textit{indirect liability rule}, which assigns full liability to the first agent in the chain, does not have a counterpart in our setting, as it may violate the feasibility constraints. But we could consider instead in our setting the rule that assigns all the flow to the most downstream agent along the river (as mentioned in the previous paragraph).  Finally, \cite{Gudmundsson2024} characterize the so-called \textit{fixed fraction rules}, which are simply obtained via the convex combinations between their \textit{direct liability rule} and \textit{indirect liability rule}.\footnote{The so-called \textit{externality-adjusted proportional rules} in \cite{yang2023pollute} adjust these rules to the so-called \textit{river pollution claims problem}.} As mentioned above, our compromise rules are obtained via the convex combinations between the no-transfer rule and the egalitarian full transfer rule.

\section*{Appendix. Proof of the results}

We start with the proof of Theorem \ref{thm1}, our most general result.

\subsection*{Proof of Theorem \ref{thm1}}
\begin{proof}
We prove first the straightforward implications of the statement. Notice that the family of rules can be formulated as follows. For each $i \in N$,
$$
R^\alpha_i(e)=\alpha_i e_i + \sum_{k < i} \frac{(1-\alpha_k)e_k}{n-k},
$$
where $\alpha=(\alpha_1,\ldots,\alpha_n) \in [0,1]^{n-1} \times \{1\}$.
\begin{itemize}
    \item[(a)] \emph{Scale invariance}. Let $\gamma \in \mathbb{R}_+$. For each $e \in \mathcal{D}$ and each $i \in N$,
    $$
    R^\alpha_i(\gamma e)=\alpha_i \gamma e_i + \sum_{k < i} \frac{(1-\alpha_k) \gamma e_k}{n-k} = \gamma R^\alpha_i(e).
    $$
    \item[(b)] \emph{Downstream impartiality}. Let $e,e' \in \mathcal{D}$, such that $e_i<e'_i$ for some $i \in N$, and $e'_j = e_j$ for all $j \in N \backslash \{i\}$,  if $k,l>i$ are such that $e_k=e_l$, we have that
    \begin{align*}
    R^\alpha_k(e') - R^\alpha_k(e) &= \alpha_k e'_k + \sum_{j < k} \frac{(1-\alpha_j)e'_j}{n-j} -  \alpha_k e_k - \sum_{j < k} \frac{(1-\alpha_j)e_j}{n-j}\\
    &= \frac{(1-\alpha_i)e'_i}{n-i} + \sum_{\substack{j<k \\ j \neq i}} \frac{(1-\alpha_j)e'_j}{n-j} - \frac{(1-\alpha_i)e_i}{n-i} - \sum_{\substack{j<k \\ j \neq i}} \frac{(1-\alpha_j)e_j}{n-j} \\
    &= \frac{(1-\alpha_i)(e'_i-e_i)}{n-i}.
    \end{align*}
    Analogously, 
    \begin{align*}
    R^\alpha_l(e') - R^\alpha_l(e) &= \frac{(1-\alpha_i)(e'_i-e_i)}{n-i}.
    \end{align*}
    Then, $R^\alpha_k(e') - R^\alpha_k(e) = R^\alpha_l(e') - R^\alpha_l(e)$ and the axiom holds. 
    \item[(c)] \emph{Upstream invariance}. Let $e,e' \in \mathcal{D}$, such that $e_i<e'_i$ for some $i \in N$, and $e_j = e'_j$ for all $j \in N \backslash \{i\}$, for each $k<i$ we have that
    $$
    R^\alpha_k(e') = \alpha_k e'_k + \sum_{j < k} \frac{(1-\alpha_j)e'_j}{n-j} = \alpha_k e_k + \sum_{j < k} \frac{(1-\alpha_j)e_j}{n-j}=R^\alpha_k(e).
    $$
\end{itemize}
We now focus on the converse implication. Let $R$ be a rule satisfying all these axioms. Let $e \in \mathcal{D}$. We proceed by induction on the number of non-null inflows in $e$. Notice that 
$R(0,\ldots,0)=(0,\ldots,0)$. Let us define $e^1=(e_1,0,\ldots,0) \in \mathcal{D}$. By \emph{downstream impartiality}, 
$$
R_j(e^1)-R_j(0,\ldots,0)=R_k(e^1)-R_k(0,\ldots,0),
$$
for each pair $j,k>1$. As $R_j(0,\ldots,0)=R_k(0,\ldots,0)=0$ for each pair $j,k>1$, we obtain that $R_j(e^1)=R_k(e^1)$ for each pair $j,k>1$. Therefore, there exists $\alpha_1(e_1) \in [0,1]$ such that 
$$
R(e^1) = \left( \alpha_1(e_1) e_1, \frac{(1-\alpha_1(e_1))e_1}{n-1}, \ldots, \frac{(1-\alpha_1(e_1))e_1}{n-1} \right).
$$
We show that $\alpha_1(e_1)$ does not depend on $e_1$ (i.e., $\alpha_1(e_1)=\alpha_1$ for each $e_1 \in \mathbb{R}_{++}$). Indeed, let $\overline{e}^1 =(\overline{e}_1,0,\ldots,0)$ with $\overline{e}_1 \neq e_1$. \emph{Scale invariance} implies that $R(\overline{e}^1) = \frac{\overline{e}_1}{e_1} R(e^1)$, and then $\alpha_1(\overline{e}_1) \overline{e}_1 = \frac{\overline{e}^1}{e^1} \alpha_1(e_1) e_1$, from where we conclude that $\alpha_1(e_1) = \alpha_1(\overline{e}_1)$ for each pair $e_1, \overline{e}_1 \in \mathbb{R}_{++}$.

Let $e^2=(e_1,e_2,0,\ldots,0) \in \mathcal{D}$. By \emph{upstream invariance}, $R_1(e^2)=R_1(e^1)=\alpha_1 e_1$. By \emph{downstream impartiality}, 
$$
R_j(e^2)-R_j(e^1)= R_k(e^2)-R_k(e^1), 
$$
for each pair $j,k>2$. 
As $R_j(e^1)=R_k(e^1)$ for each pair $j,k>2$, we obtain that $R_j(e^2)=R_k(e^2)$, for each pair $j,k>2$. Thus, there exists $\alpha_2(e_2) \in [0,1]$ such that $R_2(e^2)=\alpha_2(e_2) e_2 + \frac{(1-\alpha_1)e_1}{n-1}$. Then,
\begin{align*}
R(e^2) =& \left( \alpha_1 e_1, \alpha_2(e_2) e_2 + \frac{(1-\alpha_1)e_1}{n-1}, \frac{(1-\alpha_1)e_1}{n-1} + \frac{(1-\alpha_2(e_2))e_2}{n-2}, \ldots, \right. \\[0.2cm]
& \left. \frac{(1-\alpha_1)e_1}{n-1} + \frac{(1-\alpha_2(e_2))e_2}{n-2} \right).
\end{align*}

We show that $\alpha_2(e_2)=\alpha_2$ for each $e_2 \in \mathbb{R}_{+}$. Indeed, let $\overline{e}^2 =(e_1,\overline{e}_2,0,\ldots,0)$ with $\overline{e}_2 \neq e_2$. By \emph{scale invariance}, $R(\overline{e}^2) = \frac{\overline{e}_2}{e_2} R\left( \frac{e_2}{\overline{e}_2} e_1, e_2, 0, \ldots, 0 \right)$, and then $\alpha_2(\overline{e}_2) \overline{e}_2 + \frac{(1-\alpha_1)e_1}{n-1} = \frac{\overline{e}_2}{e_2} \left[ \alpha_2(e_2) e_2 + \frac{(1-\alpha_1) \frac{e_2}{\overline{e}_2} e_1}{n-1} \right]$, from where we conclude that $\alpha_2(e_2) = \alpha_2(\overline{e}_2)$ for each pair $e_2, \overline{e}_2 \in \mathbb{R}_{+}$.

Assume now that the proof holds for $e^i=(e_1,e_2,\ldots,e_i,0,\dots,0)$. We show that it is also true for $e^{i+1}=(e_1,e_2,\ldots,e_i,e_{i+1},0\dots,0)$. By \emph{upstream invariance}, $R_j(e^{i+1})=R_j(e^i)$ for each $j < i+1$. By \emph{downstream impartiality}, 
$$
R_j(e^{i+1})-R_j(e^i)= R_k(e^{i+1})-R_k(e^i), 
$$
for each pair $j,k>i+1$. As $R_j(e^i)=R_k(e^i)$ for each pair $j,k>i+1$, we obtain that $R_j(e^{i+1})=R_k(e^{i+1})$ for each pair $j,k>i+1$. Thus, there exists $\alpha_{i+1}(e_{i+1}) \in [0,1]$ such that $R_{i+1}(e^{i+1})=\alpha_{i+1}(e_{i+1}) e_{i+1} + \sum_{k<i+1} \frac{(1-\alpha_k)e_k}{n-k}$. Then, applying the induction hypothesis,
\begin{align*}
R(e^{i+1}) =& \left( \alpha_1 e_1, \ldots, \alpha_{i+1}(e_{i+1}) e_{i+1} + \sum_{k<i+1} \frac{(1-\alpha_k)e_k}{n-k}, \ldots, \right. \\[0.2cm]
& \left.\sum_{k< i+1} \frac{(1-\alpha_k)e_k}{n-k} + \frac{(1-\alpha_{i+1}(e_{i+1}))e_{i+1}}{n-(i+1)} \right).
\end{align*}
We show that $\alpha_{i+1}(e_{i+1})=\alpha_{i+1}$ for each $e_{i+1} \in \mathbb{R}_{+}$. Indeed, let $\overline{e}^{i+1} =(e_1,\ldots, e_i,\overline{e}_{i+1},0,\ldots,0)$ with $\overline{e}_{i+1} \neq e_{i+1}$. By \emph{scale invariance}, $R(\overline{e}^{i+1}) = \frac{\overline{e}_{i+1}}{e_{i+1}} R\left( \frac{e_{i+1}}{\overline{e}_{i+1}} e_1, \ldots, \frac{e_{i+1}}{\overline{e}_{i+1}} e_i, e_{i+1}, 0, \ldots, 0 \right)$, and then $ \alpha_{i+1}(\overline{e}_{i+1}) \overline{e}_{i+1} + \sum_{k<i+1} \frac{(1-\alpha_k)e_k}{n-k} = \frac{\overline{e}_{i+1}}{e_{i+1}} \left[ \alpha_{i+1}(e_{i+1}) e_{i+1} + \sum_{k<i+1} \frac{(1-\alpha_k) \frac{e_{i+1}}{\overline{e}_{i+1}} e_k}{n-k} \right]$, from where we conclude that $\alpha_{i+1}(e_{i+1}) = \alpha_{i+1}(\overline{e}_{i+1})$ for each $e_{i+1}, \overline{e}_{i+1} \in \mathbb{R}_{+}$.
\end{proof}

We can now provide the proofs for the remaining results.

\subsection*{Proof of Theorem \ref{ShTh}}
\begin{proof}
We prove first that the Shapley rule satisfies the properties in the statement. Theorem \ref{thm1} guarantees that it fulfills \emph{scale invariance}, \emph{downstream impartiality}, and \emph{upstream invariance}. As for \emph{balance}, let $e \in \mathcal{D}$, such that $e_i>0$ for some $i \in \{1,\dots, n-1\}$, and $e_j=0$ for all $j \in N \backslash \{i\}$. Then,
$$
R^{Sh}_k(e) = 
\begin{cases}
0 & \text{if } k<i \\
\dfrac{e_i}{n-i+1} & \text{if } k \geq i.
\end{cases}
$$
Therefore,
$$
\frac{1}{n-i}\sum_{k>i} R^{Sh}_k(e) = \frac{1}{n-i}\sum_{k>i} \frac{e_i}{n-i+1} = \frac{1}{n-i} \frac{e_i}{n-i+1} (n-i) =  \frac{e_i}{n-i+1} = R^{Sh}_i(e).
$$
We now focus on the converse implication. Let $R$ be a rule satisfying all these axioms. Let $e \in \mathcal{D}$. By Theorem \ref{thm1}, there exists $(\alpha_1,\ldots,\alpha_{n-1}) \in [0,1]^{n-1}$ such that 
    \begin{align*}
R(e) =& \left( \alpha_1 e_1, \alpha_2 e_2+\frac{e_1\left(1-\alpha_1\right)}{n-1}, \alpha_3 e_3+\frac{e_1\left(1-\alpha_1\right)}{n-1}+\frac{e_2\left(1-\alpha_2\right)}{n-2}, \ldots, \right.\\
    & \left. \alpha_n e_n+\frac{e_1(1-\alpha_1)}{n-1}+\ldots+\frac{e_{n-1}\left(1-\alpha_{n-1}\right)}{n-(n-1)} \right),
\end{align*}
with $\alpha_n=1$. Now, for each $i \in \{1,\dots, n-1\}$, 
$$
\alpha_i = R_i(\overbrace{0, \ldots, 0}^{i-1}, 1, \overbrace{0, \ldots, 0}^{n-i}).
$$
By \emph{balance}, 
$$
\alpha_i = \frac{1}{n-i}\sum_{k>i} R_k(\overbrace{0, \ldots, 0}^{i-1}, 1, \overbrace{0, \ldots, 0}^{n-i}) = \frac{1}{n-i} \frac{1-\alpha_i}{n-i} (n-i)=\frac{1-\alpha_i}{n-i}.
$$
That is, $\alpha_i = \frac{1}{n-i+1}$ for any $i \in \{1,\ldots, n-1\}$, which implies that $R$ and $R^{Sh}$ coincide.
\end{proof}

\subsection*{Proof of Theorem \ref{FamilyTh}}
\begin{proof}
We prove first that any compromise rule satisfies the properties in the statement. As a compromise rule, $R^\lambda$ is a special case of the family described in Theorem \ref{thm1}.  It clearly satisfies \emph{scale invariance}, \emph{downstream impartiality}, and \emph{upstream invariance}. We then consider the remaining axiom, \emph{equal treatment of equal source inflows}. Let $e,e' \in \mathcal{D}$ such that $e_{s(e)}=e'_{s(e')}$, then 
$$
R^\lambda_{s(e)}(e) =  \lambda e_{s(e)} =  \lambda e'_{s(e')} = R^\lambda_{s(e')}(e').
$$
Conversely, let $R$ be a rule satisfying the four axioms. Let $e \in \mathcal{D}$. By Theorem \ref{thm1}, there exists $(\alpha_1,\ldots,\alpha_{n-1}) \in [0,1]^{n-1}$ such that 
\begin{align*}
R(e) =& \left( \alpha_1 e_1, \alpha_2 e_2+\frac{e_1\left(1-\alpha_1\right)}{n-1}, \alpha_3 e_3+\frac{e_1\left(1-\alpha_1\right)}{n-1}+\frac{e_2\left(1-\alpha_2\right)}{n-2}, \ldots, \right.\\
    & \left. \alpha_n e_n+\frac{e_1(1-\alpha_1)}{n-1}+\ldots+\frac{e_{n-1}\left(1-\alpha_{n-1}\right)}{n-(n-1)} \right),
\end{align*}
with $\alpha_n=1$. Now, for each $i \in \{1,\dots, n-1\}$, 
$$
\alpha_i = R_i(\overbrace{0, \ldots, 0}^{i-1}, 1, \overbrace{0, \ldots, 0}^{n-i}).
$$
\emph{Equal treatment of equal source inflows} implies that, for each $\{k,j\} \subseteq N$,  $\alpha_k=\alpha_j=\lambda \in [0,1]$. Thus, the above becomes
    $$
    R_i(e)= \lambda e_i + \sum_{k<i} \frac{(1-\lambda)e_k}{n-k} = \lambda R_i^{NT}(e) + (1-\lambda)R_i^{EFT}(e),
    $$
for each $i \in \{1,\dots, n-1\}$, as desired.
    \end{proof}

\subsection*{Proof of Theorem \ref{NTTh}}

\begin{proof}
We prove first that the the no-transfer rule satisfies the properties in the statement. As this rule is a special case of the family described in Theorem \ref{thm1} ($\alpha_1=\ldots=\alpha_{n-1}=1$), it clearly satisfies \emph{scale invariance}, \emph{downstream impartiality}, and \emph{upstream invariance}. As for \emph{order preservation}, let $e \in \mathcal{D}$, and $i,j\in N$ such that $i<j$ and $e_i \geq e_j$. Then, it is obvious that $R^{NT}_i(e)=e_i \geq e_j =R^{NT}_j(e)$.

Conversely, let $R$ be a rule that satisfies the four axioms in the statement. Then, by Theorem \ref{thm1}, there exists a vector $\alpha=(\alpha_1, \ldots, \alpha_{n-1},\alpha_n) \in [0,1]^{n-1}\times\{1\}$ such that, for each $e \in \mathcal{D}$ and each $i\in\{i,\ldots,n-1\}$,
    $$
    R^\alpha_i(e)=\alpha_i e_i + \sum_{k<i} \frac{(1-\alpha_k)e_k}{n-k}.
    $$
    We proceed by contradiction to show that $\alpha_2=\ldots=\alpha_{n-1}=1$. Assume that there exists $\alpha_i \in \{2,\ldots,n-1\}$ such that $\alpha_i<1$. Let us consider a problem $e^i \in \mathcal{D}$ such that $e^i_1=1$, $e^i_i>\frac{n-i}{1-\alpha_i}$ and $e^i_j=0$ for each $j \in N\backslash \{1,i\}$. As $1<n$ and $e^i_1>e^i_n$, \emph{order preservation} implies that $R^\alpha_1(e^i) \geq R^\alpha_n(e^i)$. That is, for each $i \in \{2,\ldots,n-1\}$,
    $$
    \alpha_1 \ge \frac{1-\alpha_1}{n-1}+\frac{\left(1-\alpha_i\right)}{n-i} e^i_i. 
    $$
    Equivalently, 
    $$
    e^i_i \leq \frac{n \alpha_1-1}{n-1} \cdot \frac{n-i}{1-\alpha_i}.
    $$
    However, $e^i_i>\frac{n-i}{1-\alpha_i}>\frac{n \alpha_1-1}{n-1} \cdot \frac{n-i}{1-\alpha_i}$, which is a contradiction with the previous condition. Therefore,  $\alpha_2=\ldots=\alpha_{n-1}=1$. As for $\alpha_1$, notice that
    $$
    R^\alpha_1(1,\dots,1)= \alpha_1, \quad \text{ and } \quad R^\alpha_2(1,\dots,1)= 1 + \frac{1-\alpha_1}{n-1}.
    $$
    \emph{Order preservation} requires that $R^\alpha_1(1,\dots,1) \geq R^\alpha_2(1,\dots,1)$, which implies that $\alpha_1 \ge 1$, and therefore $\alpha_1=1$.
    \end{proof}

\subsection*{Proof of Theorem \ref{thm6}}
    \begin{proof}
We prove first that any partial compromise rule satisfies the properties in the statement. As a partial compromise rule $R^\delta$ is a special case of the family described in Theorem \ref{thm1} ($\alpha_1=\delta$ and $\alpha_i=1-\frac{n-i}{n-1}(1-\delta)$ for $i \in \{2,\ldots,n-1\}$), it clearly satisfies \emph{scale invariance}, \emph{downstream impartiality}, and \emph{upstream invariance}. As for \emph{equal treatment of equal upstream total inflow}, let $e,e' \in \mathcal{D}$ such that there exists $i \in N$ with $e_i=e'_i$, and $\sum_{j<i} e_j = \sum_{j<i} e'_j$. Then, 
\begin{align*}
R^\delta_i(e) &= \delta e_i + (1-\delta)\left[ \left( 1-\frac{n-i}{n-1} e_i \right) + \frac{1}{n-1} \sum_{j<i}e_j \right] \\
&= \delta e'_i + (1-\delta)\left[ \left( 1-\frac{n-i}{n-1} e'_i \right) + \frac{1}{n-1} \sum_{j<i}e'_j \right] \\
&=R^\delta_i(e').
\end{align*}

Conversely, let $R$ be a rule satisfying the four axioms. Then, by Theorem \ref{thm1}, there exists a vector $\alpha=(\alpha_1, \ldots, \alpha_{n-1},\alpha_n) \in [0,1]^{n-1}\times\{1\}$ such that, for each $e \in \mathcal{D}$ and each $i\in\{i,\ldots,n-1\}$,
        $$
        R^\alpha_i(e)=\alpha_i e_i + \sum_{k<i} \frac{(1-\alpha_k)e_k}{n-k}.
        $$
        Notice that
        $$
        R^\alpha_n(1_1,0_{-1})= \frac{1-\alpha_1}{n-1},
        $$
        and, for each $i \in \{2,\ldots,n-1\}$,
        $$
        R^\alpha_n(1_i,0_{-i})= \frac{1-\alpha_i}{n-i}.
        $$
        As $\sum_{j<n}(1_1,0_{-1}) = \sum_{j<n}(1_i,0_{-i})$ for each $i \in \{2,\ldots,n-1\}$, \emph{equal treatment of equal upstream total inflow} implies that, for each $i \in \{2,\ldots,n-1\}$,
        $$
        R^\alpha_n(1_1,0_{-1})=R^\alpha_n(1_i,0_{-i}).
        $$ 
        Therefore, $\frac{1-\alpha_i}{n-i}=\frac{1-\alpha_1}{n-1}$, or equivalently, $\alpha_i = 1-\frac{\left(1-\alpha_1\right)}{n-1}(n-i)$ for each $ i \in \{2,\ldots,n-1\}$. Substituting in the general expression of $R^\alpha$ we obtain that 
        $$
        R^\alpha_i(e) = \left[1-\frac{1-\alpha_1}{n-1}(n-i)\right] e_i+\frac{1-\alpha_1}{n-1} \sum_{k<i} e_k.
        $$
        Let $\delta=\alpha_1$. Then, 
        \begin{align*}
        \delta R^{NT}_i(e) + (1-\delta)R^{EPT}_i(e) &= \delta e_i + (1-\delta) \left[ \left[1-\frac{(n-i)}{n-1}\right] e_i + \frac{1}{n-1} \sum_{k<i} e_k \right] \\
        &= \left[1-\frac{1-\delta}{n-1}(n-i)\right] e_i+\frac{1-\delta}{n-1} \sum_{k<i} e_k \\
        &= R^\alpha_i(e).
        \end{align*}
        Finally, as $\alpha_1 \in [0,1]$, we also have that $\delta \in [0,1]$.
        \end{proof}

\newpage

\bibliography{references.bib} 
\bibliographystyle{mystyle3}

\end{document}